\documentclass[
]{ceurart}

\usepackage{listings}
\usepackage{comment}
\usepackage{arydshln}
\begin{document}

\copyrightyear{2026}
\copyrightclause{Copyright for this paper by its authors.
  Use permitted under Creative Commons License Attribution 4.0
  International (CC BY 4.0).}

\conference{AAAI'26 Workshop (WS37), Machine Ethics: from formal methods to emergent machine ethics, January 20--27, 2026, Singapore}

\title{Learning from Mistakes: Can LLM Self-Recover after Misalignment?}


\author[1]{Olga E. Sorokoletova}[%
orcid=0009-0005-4356-2649,
email=sorokoletova@diag.uniroma1.it
]
\cormark[1]
\author[1]{Francesco Giarrusso}[%
orcid=0009-0006-9639-819X,
email=giarrusso@diag.uniroma1.it,
]
\author[1]{Vincenzo Suriani}[%
orcid=0000-0003-1199-8358,
email=suriani@diag.uniroma1.it,
]
\author[1]{Daniele Nardi}[%
orcid=0000-0001-6606-200X,
email=nardi@diag.uniroma1.it,
]
\address[1]{Sapienza University of Rome, Piazzale Aldo Moro, 5, Roma, 00185, Italy}

\cortext[1]{Corresponding author.}

\begin{abstract}
Responsible AI initiatives place great emphasis on the safety of Large Language Model (LLM)-based systems. In particular, it has become standard practice to subject these models to an alignment procedure aimed at preventing harmful outputs. However, once aligned, a model is not guaranteed to maintain this alignment throughout its lifecycle. Moreover, the likelihood of misalignment increases as malicious actors may deliberately employ jailbreaking techniques to compromise LLM safety. To counter this, much research has focused on improving alignment methods and post-processing filters. In this paper, we introduce a new perspective on advancing LLM alignment: rather than developing stronger alignment techniques, we investigate the model's intrinsic ability to recover its alignment after corruption. We propose a methodology for modeling the safety trajectories of user-assistant interactions and for detecting recovery trends within them. We apply this approach to a jailbreaking scenario, presenting a preliminary recovery analysis based on a dataset of adversarial multi-turn dialogues and examining the influence of the content moderation model chosen for safety evaluation. Project page with an interactive data visualizer is available at \href{https://lab-rococo-sapienza.github.io/LearningfromMistakes/}{https://lab-rococo-sapienza.github.io/LearningfromMistakes/}.
\end{abstract}

\begin{keywords}
  Large Language Model (LLM) \sep
  LLM Safety \sep
  Alignment \sep
  Safeguarding \sep
  Content Moderation Tools \sep
  Jailbreaking
\end{keywords}

\maketitle

\section{Introduction}
Large Language Models (LLMs) are trained on vast amounts of unprocessed, Internet-scraped data, which makes their outputs inherently associated with various risks \cite{ISRSAA2025}, such as facilitating crimes or self-harm, exposing private or sensitive information, and producing hateful or discriminatory content. To mitigate these risks, most LLMs undergo safety \textit{alignment}, meaning they are trained to avoid generating unsafe instructions and to conform to human norms and expectations in their responses. 

Although numerous efforts have been made to enhance LLM safety through improved alignment techniques \cite{ziegler2020, stiennon2020, ouyang2022, rafailov2024} and defensive mechanisms \cite{jain2023, cao2024, zeng2024, deng2024}, many researchers agree that no system can be considered perfectly aligned and that any model's safety can eventually be compromised \cite{russinovich2025}. 
This issue becomes even more critical when malicious actors deliberately attempt to compromise a model's safety through adversarial means, such as employing jailbreaking techniques \cite{russinovich2025, zeng2024, deng2024, zou2023}.

In multi-turn settings, risk is not only whether the model complies once, but whether successive turns accumulate context and progressively increase the severity or harmfulness of the interaction. Moreover, once misaligned, a model may remain exploitable across subsequent turns for additional malicious purposes. From a risk-mitigation perspective, a dynamic view of the full interaction allows us to assess whether misalignment stabilizes, escalates, or whether the model exhibits the ability to self-limit undesirable behavior.

Unlike previous works that focus on developing more robust alignment methods \cite{cao2024towards}, our study takes a different perspective. Instead of striving to design systems that never drift into misalignment, we introduce a new methodology of analysis to gain insights into \textit{self-recoverability}, thus analyzing, with a novel methodology, whether models can restore their alignment without external intervention.

We narrow our investigation to cases when misalignment occurred due to adversarial prompting attacks in multi-turn dialogues. To this end, we organized a red teaming challenge among master's students specializing in Artificial Intelligence and Robotics, during which we collected a dataset of adversarial interactions. Using this data, we propose a method for modeling dialogue safety as a trajectory and analyzing its alignment dynamics to identify self-recovery patterns. Furthermore, we discuss relevant metrics and quantitative measures for assessing a model's ability to learn from its safety failures and report findings and statistics derived from the dataset, including insights into correlations with risk categories and the choice of the safety evaluation tool.

\section{Related Works}
LLMs' safety evaluation typically relies on either specialized models (safety-specific classifiers and moderation systems), LLM-as-a-Judge approaches, or ensemble methods that combine multiple evaluation strategies~\cite{vidgen2024, russinovich2025}. The LLM-as-a-Judge method, in which different backbone models, such as those from the GPT family \cite{zheng2023,  liu2023gevalnlgevaluationusing}, are used as evaluators, is gaining increasing adoption. When properly implemented, these strategies can achieve substantial alignment with human judgment~\cite{zheng2023}. 

In evaluation pipelines, judge models are typically prompted with detailed instructions and scoring rubrics~\cite{qi2023finetuning, jindal2025sage}. However, this process remains dependent on prompt design and rubric specificity~\cite{yamauchi2025empiricalstudyllmasajudgedesign}.

In deployed systems, safety assessment is performed through moderation pipelines that employ fine-tuned classifiers rather than general-purpose prompted models~\cite{markov2023holistic, inan2023}. Practical moderation systems typically involve taxonomy design and a model-based classification targeting policy-defined harm categories~\cite{markov2023holistic}. Safety-evaluation tools can be employed at different stages of the pipeline: during evaluation, at inference time, or during training itself. AI service providers typically develop proprietary fine-tuned classifiers for these purposes, although the specific architectures, training data, and methodologies typically remain undisclosed. For instance, Anthropic's Constitutional AI approach uses models guided by a set of ethical principles to critique and revise model outputs during training~\cite{bai2022constitutional}. It deploys customized versions of their models as real-time classifiers in production systems \cite{anthropic2024safeguards}.

Beyond proprietary stacks, open evaluation models have been released that assess safety based on explicit taxonomies and are intended for turn-level screening of both prompts and responses. These specialized safety tools are typically accessible via APIs (e.g., Google Perspective API\footnote{\url{https://perspectiveapi.com/}}, Azure Content Safety API\footnote{\noindent\url{https://ai.azure.com/explore/contentsafety}}, OpenAI Content Moderation API\footnote{\url{https://platform.openai.com/docs/guides/moderation/overview}}) or as downloadable models fine-tuned for safety classification, such as Llama Guard \cite{inan2023} or Qwen Guard \cite{zhao2025}.

Most prior works treat alignment as a static property, focusing on aggregate scores or binary success/failure judgments rather than characterizing temporal alignment dynamics at the level of interaction trajectories. However, multi-turn attacks have proven highly effective~\cite{li2024llmdefensesrobustmultiturn,russinovich2025}. Under these premises, content can become progressively more problematic due to incremental jailbreaks and the accumulation of context. Identifying the exact moment when misalignment occurs can be challenging; in certain cases, jailbreaking can be understood as a progressive erosion of safety constraints. This perspective suggests that model performance should be characterized not only by error avoidance but also by the ability to return to acceptable behavior after a failure.

To the best of our knowledge, the question of whether and to what extent an LLM in a multi-turn setting can re-establish alignment after a safety failure has not been systematically studied as a primary objective.
Motivated by this gap, our contribution lies in proposing a trajectory-centric methodology to represent safety over time and to gain insights into recovery-relevant structural patterns.

To model LLM self-recovery within a dialogue, it is necessary to assign a safety flag to each user query and model response at every interaction turn. Additionally, since we aim for an analysis that accounts for the risk category of prompts and responses, the evaluation system must provide a multi-class classification consistent with the chosen risk taxonomy.

In this work, we adopt \textit{Llama Guard}, a Llama-based model fine-tuned for content safety classification. The model covers the risk categories listed in \autoref{tab:cats}.
Using Llama Guard represents a deliberate design choice in this study, as our goal is not to identify the most effective safeguarding model but to propose a methodology for modeling alignment dynamics through one such tool. 

\section{Dataset}
We conducted a red-teaming challenge involving 48 master's students enrolled in the course ``Seminars in Artificial Intelligence and Robotics'' of Sapienza University of Rome, to collect a dataset of adversarial dialogues. The challenge comprised multiple jailbreaking tasks spanning different objectives. However, not all tasks were compatible with the safety risk taxonomy used by Llama Guard, which is inherited from the MLCommons taxonomy\footnote{\url{https://mlcommons.org/2024/04/mlc-aisafety-v0-5-poc/}}. In particular, several tasks targeted infrastructural aspects, such as extracting system prompt information, and hence did not correspond to any defined hazard category. 

Moreover, some conversations are not independent samples but branching trajectories originating from a common initial prompt where users iteratively refine or slightly modify an attack to exploit model stochasticity. This process creates conversations that exhibit a degree of similarity and share a common root. When multiple branches correspond to the same attack attempt, we retained only one representative instance, specifically the longest and most refined version of the attack.

As a result, we selected 597 conversations out of the original 1364 for analysis. The retained subset contains 7635 messages, corresponding to 2454 user-assistant turns, and covers a range of safety-relevant objectives, including the elicitation of gender and racial bias, privacy violations, promotion of physical and non-physical harm, and the induction of hallucinations. 

During the challenge, participants were provided with a black-box interface for querying the model and a two-hour time window. The model under attack was \texttt{Minerva-7B-instruct-v1.0} \cite{orlando-etal-2024-minerva}, trained on nearly 2.5 trillion tokens. 
The challenge was conducted primarily in Italian, except in cases where a jailbreaking technique was language-dependent (e.g., multilingual attacks \cite{deng2024}) or when a participant was a non-Italian speaker. 

The dialogues were multi-turn, but all types of adversarial prompting attacks were permitted. These included single-turn and multi-turn attacks, single-shot and many-shot strategies, isolated and combined techniques, human-crafted prompts, as well as optimization-based prompts that are transferable across models. This open setting was intended to capture a broad spectrum of real-world behaviors.

Each user-assistant interaction was manually annotated both for the success or failure of the attack and with multi-class labels indicating the specific jailbreak techniques employed. For this purpose, we relied on a comprehensive taxonomy of jailbreaking attacks, developed on the basis of attacks delivered during the challenge and introduced in \cite{sorokoletova2025guarding}. This taxonomy spans three hierarchical levels and groups 50 attack techniques into seven distinct families, defined by the primary mechanism through which adversarial prompts bypass safeguards: \textit{Impersonation Attacks \& Fictional Scenarios}, \textit{Privilege Escalation}, \textit{Persuasion}, \textit{Cognitive Overload \& Attention Misalignment}, \textit{Encoding \& Obfuscation}, \textit{Goal-Conflicting Attacks}, and \textit{Data Poisoning Attacks}. While this fine-grained annotation enables detailed analysis of attack strategies, it is not the primary focus of the present work.

To assess the reliability of the automatic safety evaluation used throughout our analysis, we measured the agreement between the safety flags assigned by the Llama Guard evaluator and our \textit{ground truth}, defined as manual annotations produced by the challenge organizers. For Llama Guard 3-1B, we observed an overall agreement of 77.08\%. Among the matching judgments, 83.91\% correspond to interactions labeled as safe by both Llama Guard and the manual annotations, while 16.09\% correspond to interactions labeled as unsafe by both. For Llama Guard 3-8B, the agreement is lower at 69.54\%, with 84.71\% of matching cases labeled as safe and 15.29\% labeled as unsafe. While these agreement levels are not ideal, they should be interpreted in the context of the task and labeling setup. Automatic safety classification is inherently noisy, and imperfect agreement is a known and widely documented property of LLM-based safety evaluators.

An additional source of disagreement arises from a mismatch in labeling granularity. We assigned the safety outcome at the dialogue level, labeling a conversation as safe or unsafe based on whether the overall jailbreak attempt was judged successful. In contrast, Llama Guard produces safety labels at the turn level. To align the two approaches, we treat a conversation as safe under Llama Guard only if all model responses across its turns are labeled safe.

The goal of the present work is not to provide definitive quantitative estimates of recovery behavior, but to introduce a novel methodology for modeling safety trajectories and to study preliminary self-recovery patterns under adversarial prompting. Within this exploratory setting, the observed agreement levels are sufficient to support qualitative and aggregate analyses. Nevertheless, improving the fidelity of safety labels remains an important direction for future work. 

\begin{figure}[t]
    \centering
    \includegraphics[width=\linewidth]{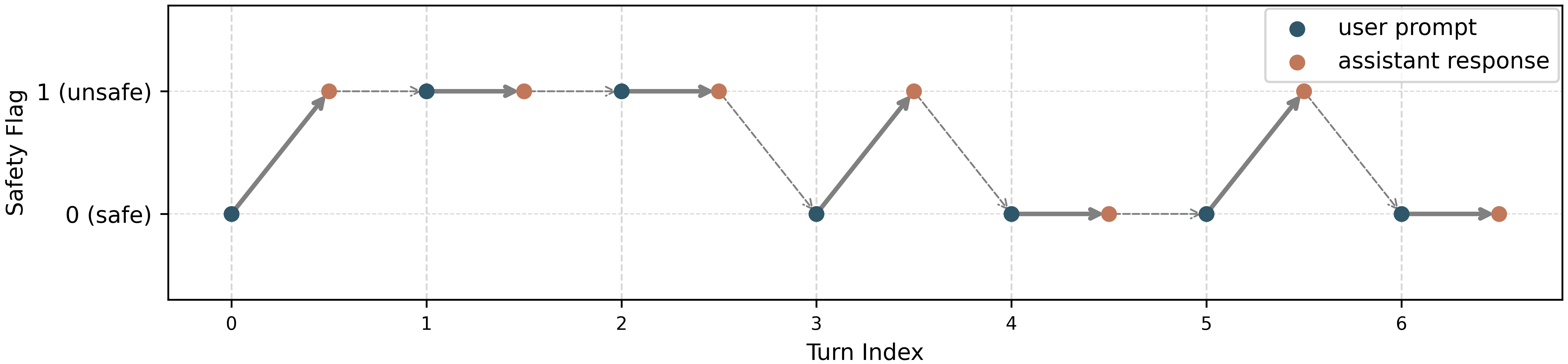}
    \caption{Safety trajectory for one of the conversations of the collected dataset. The first misalignment occurs at turn 0, followed by recovery at turn 4. The second misalignment occurs at turn 5, with recovery at turn 6. Misalignment-recovery pairs are identified by transitions from unsafe model responses (flagged as 1) to the first subsequent safe response (flagged as 0).}
    \label{fig:traj}
\end{figure}

\section{Modeling Alignment Recovery}
Analyzing the temporal safety dynamics of user–LLM interactions can help to explain incremental jailbreaks better and to understand how models recover after different types of safety failures. By examining the evolution of safety states throughout a dialogue, we can trace the path from alignment to misalignment and, in some cases, to self-recovery. To capture this process, we first introduce the concept of \textit{safety trajectories}, which model the sequential behavior of safety signals across interaction turns, and then identify \textit{recovery trends} that describe how and when the model regains alignment after a failure event.

\subsection{Safety Trajectories}
Once a trusted content moderation tool is chosen and safety flags are assigned, the safety trajectory is obtained by plotting these values for each user prompt and assistant response on the vertical axis against the corresponding dialogue turn on the horizontal axis, as shown in \autoref{fig:traj}. When turn-level ground truth safety labels are available, an analogous trajectory can be constructed directly from those labels.

Bold arrows in \autoref{fig:traj} represent the trend at each interaction turn. 
A horizontal arrow at y = 0 indicates a benign exchange (turns 4 and 6), while a horizontal arrow at y = 1 indicates a successful attack (turns 1 and 2). An ascending arrow corresponds to broken alignment, where the model gives an unsafe response to a benign prompt (turns 0, 3, and 5). Finally, a descending arrow would represent resistance to an attack at that turn (not present in \autoref{fig:traj}).

To provide more context, \autoref{fig:traj} illustrates the safety trajectory of a conversation in which the user attempted to elicit racial bias by repeatedly applying a predefined multi-technique jailbreak prompt known to be effective in practice, commonly referred to as the \textit{ChatGPT Khajiit Jailbreak Prompt}\footnote{\href{https://deepleaps.com/prompt/chatgpt-khajiit-jailbreak-prompt/}{ChatGPT Khajiit Jailbreak Prompt}}. This prompt combines elements of role-playing and fictional framing to bypass safety constraints and has circulated widely in practitioner and user communities. In this interaction, the user instructed the model to generate a tabular mapping from racial groups to insults, corresponding to the ``Hate'' risk category in the Llama Guard taxonomy. Consistently, all unsafe responses produced during this dialogue were classified under this category by the evaluator.

This example highlights the importance of selecting the appropriate content moderation model for interpreting interaction dynamics. Let us consider the first 4 turns in \autoref{fig:traj}. The model drifted into misalignment immediately at turn 0, even though the user prompt was benign. The attack then began, and misalignment persisted up to turn 3, where the prompt was again innocuous. However, if the safety flags assigned by Llama Guard to user prompts at turns 0 and 3 were false negatives, these 4 turns would instead represent a multi-turn attack that became effective immediately and remained active until the user ceased the attack at turn 4.

\begin{figure}[t]
    \centering
    \includegraphics[width=0.6\linewidth]{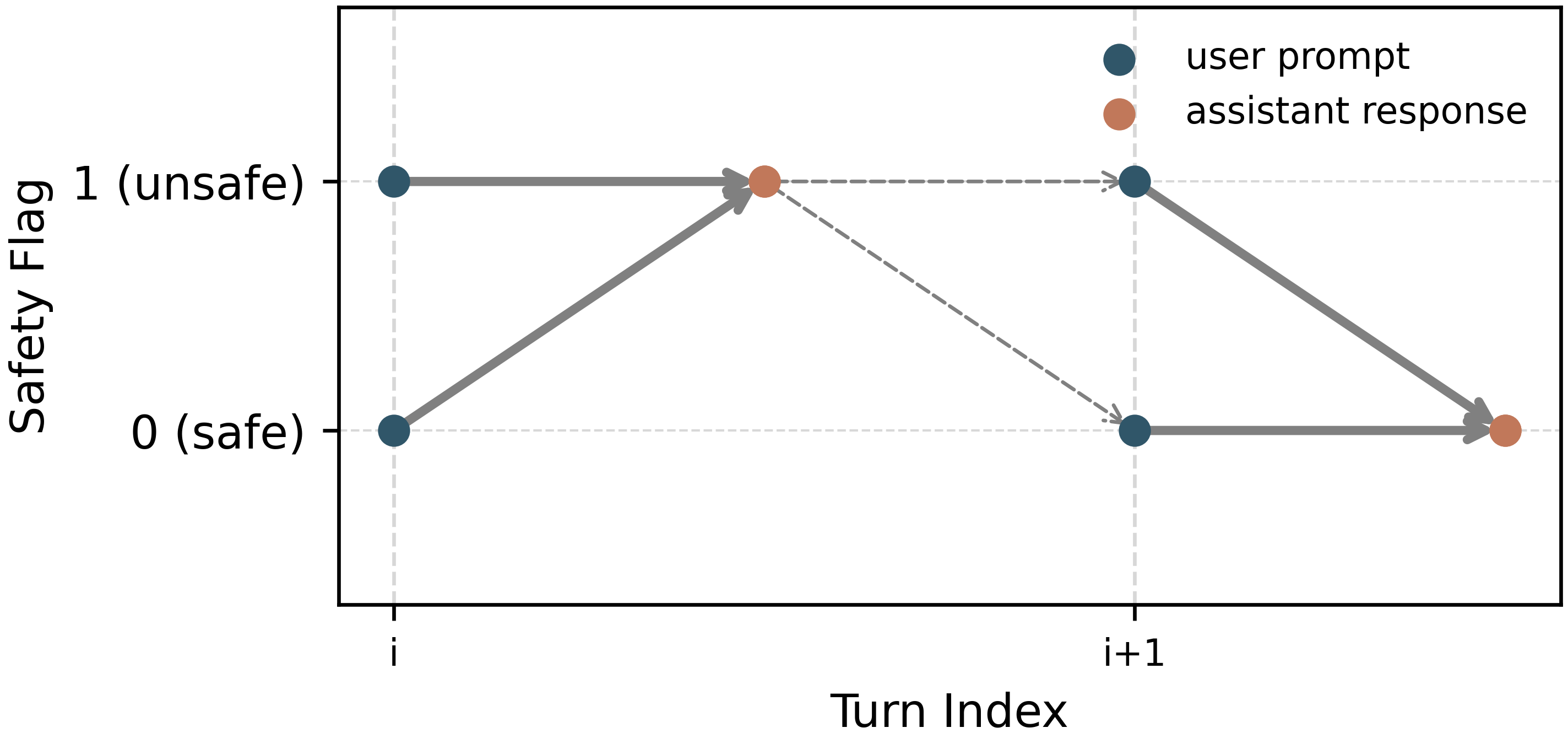}
    \caption{Schematic illustration of one-step recovery. The two possible recovery paths are shown, depending on whether the user prompt at turn $i + 1$ is safe (lower recovery path) or unsafe (upper recovery path).}
    \label{fig:rec}
\end{figure}

\subsection{Recovery Trends}
To define recovery on a safety trajectory, we first fix the turn at which an unsafe model response occurs. \textit{Recovery} is then observed when the model produces the first safe response after this misalignment event. 

For illustration, consider the case in which recovery occurs immediately in the next turn, as shown in \autoref{fig:rec}. Here, we fix $i$ as the turn where the model response is unsafe. At $i + 1$, the safety of the user prompt may vary, which is represented by branching edges, while the model response becomes safe. This yields two possible recovery paths, depending on whether the user prompt at the recovery turn is safe or unsafe: the upper recovery path and the lower recovery path. These paths capture distinct interaction dynamics that are not visible when considering recovery as a single scalar event. 

When recovery occurs after $N$ turns, the same recovery paths appear later in the trajectory. For example, in \autoref{fig:traj}, the failure at turn 0 is resolved at turn 4 along the lower recovery path. Importantly, the recovery path specifies \textit{how} the model returns to alignment, not only \textit{when} this occurs. 

Multiple recovery events may occur within a single safety trajectory. For this reason, we distinguish between absolute and temporary recovery. \textit{Absolute recovery} refers to a restoration of alignment that is maintained until the end of the interaction (e.g., the recovery at turn 6 in \autoref{fig:traj}), whereas \textit{temporary recovery} denotes a return to alignment that is later followed by a new misalignment. Repeated misalignment may arise for different reasons, including changes in the user's attack strategy or insufficiently robust recovery that fails to stabilize alignment.

The purpose of modeling recovery trends and paths is to define the structural patterns that recovery metrics aim to capture. Metrics such as \textit{Misalignment Length} and \textit{Recovery Duration} are meaningful only once recovery events are identified and contextualized within a trajectory. Recovery trends determine which misalignment-recovery pairs are counted, how overlapping events are interpreted, and how robustness is distinguished from short-term stabilization. 

\section{Analysis}

In this section, we analyze alignment dynamics over complete interaction trajectories. We first introduce the statistics used to characterize recovery behavior and misalignment over time. We then present a comparative study to assess whether these statistics change when a larger content moderation model is employed. Finally, we examine how alignment dynamics vary across different risk categories, highlighting risk-specific recovery patterns.

\begin{table}[t]
\centering
\caption{Preliminary alignment statistics showing how many conversations contain unsafe outputs and recovery events.}
\setlength{\tabcolsep}{1mm}
\begin{tabular}{lrrr}
\toprule
\textbf{Guard Model} & \shortstack{\textbf{Unsafe}\\\textbf{Conversations}} & \shortstack{\textbf{Conversations}\\\textbf{w/ Recoveries}} & \shortstack{\textbf{Conversations}\\\textbf{w/ Multiple Recoveries}} \\
\midrule
Llama Guard 3-1B  & 202 (33.8\%) & 87 (14.6\%) & 19 (3.2\%) \\
Llama Guard 3-8B  & 193 (32.3\%) & 66 (11.1\%) & 5 (0.8\%) \\
\bottomrule
\end{tabular}
\label{tab:all}
\end{table}

\begin{table}[t]
\centering
\caption{Summary of average alignment dynamics in conversations that exhibit recovery, detailing typical dialogue length, recovery frequency and timing, and the persistence of misalignment and restored alignment.}
\setlength{\tabcolsep}{1mm}
\small
\resizebox{\textwidth}{!}{
\begin{tabular}{lrrrrr}
\toprule
\textbf{Guard Model}
& \shortstack{\textbf{Avg.}\\\textbf{Turns}} 
& \shortstack{\textbf{Avg.}\\\textbf{Recoveries}}
& \shortstack{\textbf{Avg.}\\\textbf{Turns b. Recovery}}
& \shortstack{\textbf{Avg.}\\\textbf{Misalignment Length}}
& \shortstack{\textbf{Avg.}\\\textbf{Recovery Duration}} \\
\midrule
Llama Guard 3-1B  & 9.8 & 1.4 & 4.3 & 1.63 & 3.69 \\
Llama Guard 3-8B  & 8.3  & 1.1 & 5.4 & 2.14 & 2.34 \\
\bottomrule
\end{tabular}}
\label{tab:rec}
\end{table}

\subsection{Quantifying Alignment Behavior}
\autoref{tab:all} reports the general dataset statistics, while \autoref{tab:rec} focuses on average statistics for conversations in which recovery occurs. As shown in \autoref{tab:all}, approximately one third (33.8\%) of conversations are flagged as unsafe by the safeguard. 
The share with at least one recovery event is much smaller (14.6\%), and only 3.2\% exhibit multiple recoveries.

From \autoref{tab:rec}, we observe that conversations exhibiting recovery contain, on average, around 10 turns, typically with 1-2 recovery events, and the first recovery usually appears near the midpoint of the dialogue. Although participants were given a two-hour window to conduct the attacks, the average conversation length remains limited. This is because the challenge did not incentivize maintaining misalignment as long as possible. Instead, participants generally switched to a new task once they considered the current attack successful. Based on this observation, we plan to explicitly encourage prolonged interaction with the model in future editions of the challenge, to support more detailed analysis of long-range safety trajectories.

Two metrics are particularly informative for characterizing alignment dynamics: Misalignment Length and Recovery Duration. \textit{Misalignment Length} captures how many turns the model remains misaligned before recovery occurs. Systems more resistant to jailbreaking should minimize this value. \textit{Recovery Duration} describes how long the model stays aligned after returning to a safe state while the attack continues. This metric is influenced by whether alignment ends because of a renewed misalignment or because the conversation concludes. When considering only cases of a renewed misalignment, higher values of Recovery Duration indicate greater robustness to sustained adversarial pressure. In our dataset, misalignment episodes last on average about 2 turns, and recovery typically persists for about 3-4 turns. 

\subsection{Model Size Comparison}
For the model size comparison, we compute all alignment statistics after re-evaluating safety using Llama Guard 3-8B. \autoref{tab:all} and \autoref{tab:rec} report the resulting metrics. As noted above, Llama Guard 3-8B shows lower agreement with the manually annotated ground truth than the 3-1B model. This is not surprising, since increased model capability does not necessarily translate into improved safety assessment \cite{ren2024safetywashing}. The reduced agreement directly affects the reliability of trajectory-based measures derived from assigned labels. For this reason, we adopt Llama Guard 3-1B as the primary baseline in our analysis and treat the 3-8B results as an auxiliary sensitivity check rather than as a competing evaluation. At the same time, comparing the two evaluations is useful for assessing how the proposed recovery metrics behave under changes in the underlying safety evaluator.

Across both evaluators, the overall pattern is consistent: unsafe behavior and recovery events occur in a substantial share of conversations, and recoveries are often temporary rather than maintained until the end of the interaction. At the same time, the two models produce different estimates of these effects. The smaller model flags more conversations as unsafe and identifies more recovery events. Moreover, the 3-8B model yields longer estimated Misalignment Length and shorter Recovery Durations on average, as shown in \autoref{tab:rec}, which together indicate weaker or less persistent recovery behavior under this evaluator. A possible explanation is that the larger model applies stricter or less reliable safety judgments, which can fragment recovery segments and reduce the apparent persistence of recovery alignment.

\begin{table}[t]
\centering
\caption{Risk-specific counts and alignment dynamics, computed using Llama Guard (LlG) 3-1B and Llama Guard (LlG) 3-8B. The dashed line separates the five categories retained for alignment dynamics analysis from the underrepresented categories excluded from comparison. A category is retained only if it exhibits at least five recovery events under both evaluators. For each recoverability metric, the two best values among the retained categories are highlighted in bold.}
\setlength{\tabcolsep}{1mm}
\small
\resizebox{\textwidth}{!}{
\begin{tabular}{l*{4}{rr}}
\toprule
\textbf{Hazard Category} 
& \multicolumn{2}{c}{\shortstack{\textbf{Conversations}\\\textbf{}\\\textbf{}}} 
& \multicolumn{2}{c}{\shortstack{\textbf{Recoveries}\\\textbf{}\\\textbf{}}} 
& \multicolumn{2}{c}{\shortstack{\textbf{Avg. Misalignment}\\\textbf{Length}}} 
& \multicolumn{2}{c}{\shortstack{\textbf{Avg. Recovery}\\\textbf{Duration}}} \\
\cmidrule(lr){2-3} \cmidrule(lr){4-5} \cmidrule(lr){6-7} \cmidrule(lr){8-9}
& \textbf{LG 3-1B} & \textbf{LG 3-8B} 
& \textbf{LG 3-1B} & \textbf{LG 3-8B} 
& \textbf{LG 3-1B} & \textbf{LG 3-8B} 
& \textbf{LG 3-1B} & \textbf{LG 3-8B} \\
\midrule
Violent Crimes            & 63 & 36 & 28 & 9  & 1.82 & \textbf{1.78} & 2.61 & \textbf{2.50} \\
Hate                      & 60 & 71 & 28 & 27 & \textbf{1.25} & 2.63 & \textbf{5.81} & 2.28 \\
Non-Violent Crimes        & 28 & 20 & 9  & 5  & \textbf{1.33} & \textbf{1.80} & 2.17 & \textbf{2.67} \\
Indiscriminate Weapons    & 20 & 31 & 6  & 8  & 2.83 & 2.25 & 2.25 & 2.20 \\
Privacy                   & 22 & 20 & 14 & 10 & 2.00 & 2.30 & \textbf{3.83} & 1.17 \\\\
\cdashline{1-9}\\
Specialized Advice        & 18 & 1  & 6  & -- & 1.17 &    
--   & 5.50 & --   \\
Intellectual Property     & 16 & 2  & 15 & 3  & 1.80 & 2.00 & 3.69 & 4.00 \\
Defamation                & 15 & 7  & 7 & 1  & 1.29 & 1.00 & 2.71 & --   \\
Sex-Related Crimes        & 11 & -- & 3  & -- & 1.00 &     --   & 4.00 & --   \\
Child Sexual Exploitation & 10  & 1  & 3  & -- & 1.67 &    
--   & 4.67   & --   \\
Suicide \& Self-Harm      & 9  & 14 & 2  & 6  & 1.00 & 1.33 & 4.00 & 2.67 \\
Elections                 & 3  & 5  & 1  & 2  & 3.00 & 1.00 & 3.00 & 4.00 \\
Sexual Content            & 1  & -- & 2  & -- & 1.50 &     --   & 1.00 & --   \\
Code Interpreter Abuse    & -- & 1  & -- & 2  & --   & 1.00 & --   & 1.50 \\
\bottomrule
\end{tabular}}
\label{tab:cats}
\end{table}

\subsection{Risk-Specific Effects}
To identify risk-specific effects, reported in \autoref{tab:cats}, we first count how many conversations contain unsafe responses in each risk category and how many recovery events occur for those categories. This allows us to assess whether a category is sufficiently represented for reliable analysis. 
In practice, representation can differ across evaluators: some categories appear frequently under the smaller content moderation model but are nearly absent under the larger one, and more rarely, the opposite occurs. For example, under Llama Guard 3-8B, ``Defamation'' exhibits only one recovery event. 

To ensure consistency across evaluations, we consider a category adequately represented only if it is not underrepresented by either model. Concretely, we retain categories that exhibit at least five recovery events under both evaluators. Based on this criterion, we select the following top-5 categories for the analysis of alignment dynamics: ``Violent Crimes,'' ``Hate,'' ``Non-Violent Crimes,'' ``Indiscriminate Weapons,'' and ``Privacy.'' Among these, ``Violent Crimes'' is the most frequent risk type. 

The comparison of the key alignment metrics is shown in \autoref{tab:cats}. According to these results, ``Violent Crimes,'' ``Non-Violent Crimes,'' and ``Hate'' exhibit the strongest recovery behavior, whereas ``Indiscriminate Weapons'' and ``Privacy'' are associated with weaker recovery. Throughout this analysis, \textit{recoverability} is defined as the combined pattern of shorter average \textit{Misalignment Length} and longer \textit{Recovery Duration}, aggregated across both content moderation models. Importantly, these trends align with the distribution of recovery events across categories.

A plausible explanation for this pattern is that categories such as ``Violent Crimes,'' ``Non-Violent Crimes,'' and ``Hate'' are governed by more explicit and well-defined policy boundaries, which may facilitate both the detection of unsafe outputs and the model's subsequent re-alignment during interaction. By contrast, ``Indiscriminate Weapons'' and ``Privacy'' violations often involve more nuanced or context-dependent constraints, including procedural details or implicit disclosures, which may be harder for the model to correct once misalignment occurs. While this interpretation is consistent with the observed patterns, further analysis on larger datasets is necessary to assess the robustness of these trends and to disentangle category-specific effects from dataset-dependent factors.

\section{Conclusion and Future Work}
This work aims to open a discussion on alignment from a new perspective. Instead of focusing solely on making models well-aligned, we suggest examining the potential for self-recovery in imperfectly aligned systems and treat it as an additional dimension of safety behavior that can be measured, understood, and potentially improved. Moreover, the study of recovery trajectories can inform the design of safer systems and help reduce repeated safety and ethically problematic failures.

The results presented in this paper are preliminary and depend on the specific setup of our challenge, including the attacked model, the selected tasks, and their correspondence to the taxonomy used for safety evaluation. To strengthen both qualitative and quantitative conclusions, future work should expand the dataset and collect longer interactions, supported by turn-level manual safety annotation. 

As a further result, we highlight the importance of reliable risk detection models for assessing the trustworthiness of the content moderation model, which is essential for accurately modeling the safety trajectory. A broader comparative analysis of different moderation systems would be valuable for future research in this direction.

Future work may also explore how to increase the \textit{Recovery Duration} and decrease the \textit{Misalignment Length} to enhance the stability of alignment once regained. 
We provide one possible framework for modeling these patterns, and we believe that many further directions remain to be explored.

\begin{acknowledgments}
This work has been carried out while Olga Sorokoletova and Francesco Giarrusso were enrolled in the Italian National Doctorate on Artificial Intelligence run by Sapienza University of Rome. We acknowledge partial financial support from PNRR MUR project PE0000013-FAIR.
\end{acknowledgments}

\section*{Declaration on Generative AI}
During the preparation of this work, the author(s) used GPT-5.2 and the Grammarly plugin in order to: Paraphrase and reword; Improve writing style; and Grammar and spelling check. After using these tool(s)/service(s), the author(s) reviewed and edited the content as needed and take(s) full responsibility for the publication’s content. 

\bibliography{sample-ceur}


\end{document}